\documentclass[a4paper,aps,twocolumn,superscriptaddress]{revtex4}

\usepackage{epsfig}
\usepackage{amsmath,amssymb,color}

\usepackage[english]{babel}

\parskip=\medskipamount

\newcommand{\eq}[1]{(\ref{#1})}
\newcommand{\fig}[1]{Fig.\ref{#1}}

\newcommand{\be}{\begin{equation}}
\newcommand{\ee}{\end{equation}}

\newcommand{\barr}{\begin{array}}
\newcommand{\earr}{\end{array}}

\newcommand{\beqn}{\begin{eqnarray}}
\newcommand{\eeqn}{\end{eqnarray}}

\newcommand{\bs}{\begin{subequations}}
\newcommand{\es}{\end{subequations}}

\newcommand{\bw}{\begin{widetext}}
\newcommand{\ew}{\end{widetext}}

\newcommand\disp{\displaystyle}

\newcommand{\la}{\left<}
\newcommand{\ra}{\right>}

\begin{document}

\title{Statistics of layered zigzags: a two--dimensional generalization of TASEP}

\author{Mikhail Tamm}
\affiliation{Physics Department, Moscow State University, 119992, Moscow, Russia}
\author{Sergei Nechaev}
\affiliation{Laboratoire de Physique Th\'eorique et Mod\`eles Statistiques, Universit\'e
Paris--Sud, 91405 Orsay Cedex, France} \affiliation{P.N. Lebedev Physical Institute of the Russian
Academy of Sciences, 119991, Moscow, Russia}
\author{Satya N. Majumdar}
\affiliation{Laboratoire de Physique Th\'eorique et Mod\`eles Statistiques, Universit\'e
Paris--Sud, 91405 Orsay Cedex, France}

\date{\today}

\begin{abstract}

A novel discrete growth model in 2+1 dimensions is presented in three equivalent formulations: i)
directed motion of zigzags on a cylinder, ii) interacting interlaced TASEP layers, and iii) growing
heap over 2D substrate with a restricted minimal local height gradient. We demonstrate that the
coarse--grained behavior of this model is described by the two--dimensional Kardar--Parisi--Zhang
equation. The coefficients of different terms in this hydrodynamic equation can be derived from the
steady state flow--density curve, the so called `fundamental' diagram. A conjecture concerning the
analytical form of this flow--density curve is presented and is verified numerically.

\bigskip


\end{abstract}

\maketitle

It is well established in the last two decades \cite{beijeren,pischke,majumdar,kk} that the
one--dimensional Kardar--Parisi--Zhang (KPZ) equation \cite{KPZ} adequately describes the
long--range dynamics of a collective motion of hopping particles on a line known as "Asymmetric
Simple Exclusion Process" (ASEP) \cite{spitzer,liggett,spohn,dhar}. Apart from many fruitful
theoretical advantages, this ASEP--to--KPZ mapping enables a fast and simple way of modelling the
KPZ dynamics. The latter is of wide interest since it appears in various contexts, (provided the
symmetry of a nonequilibrium statistical system under discussion allows for the effective
1+1--dimensional description), including, to name but a few, the models of crystal growth
\cite{RSOS}, Molecular Beam Epitaxy \cite{MBE}, Burgers' turbulence \cite{zhang,khanin},
polynuclear growth \cite{M,PS,PNG,BR,J}, ballistic deposition \cite{Mand,MRSB,KM,BMW} etc.

It is therefore an appealing idea to seek for a similar simple discrete multi--particle system
whose long--range dynamics would be governed by a two--dimensional KPZ--type equation. Lately,
several models of the desired nature, i.e. ones which combine the discreteness with the long--range
KPZ--type dynamics, were suggested \cite{wolf,prsp,borodin1,borodin2,borodin3,odor}. In this Letter
we propose another model belonging to this class, which, in our opinion, combines the advantage of
physical transparency with the flexibility of tuning the internal parameters of the model to catch
the different desired regimes both in 1+1 and in 2+1 dimensions. This model, which we call ``Zigzag
Model'', has a simple geometrical formulation.

\begin{figure}[ht]
\epsfig{file=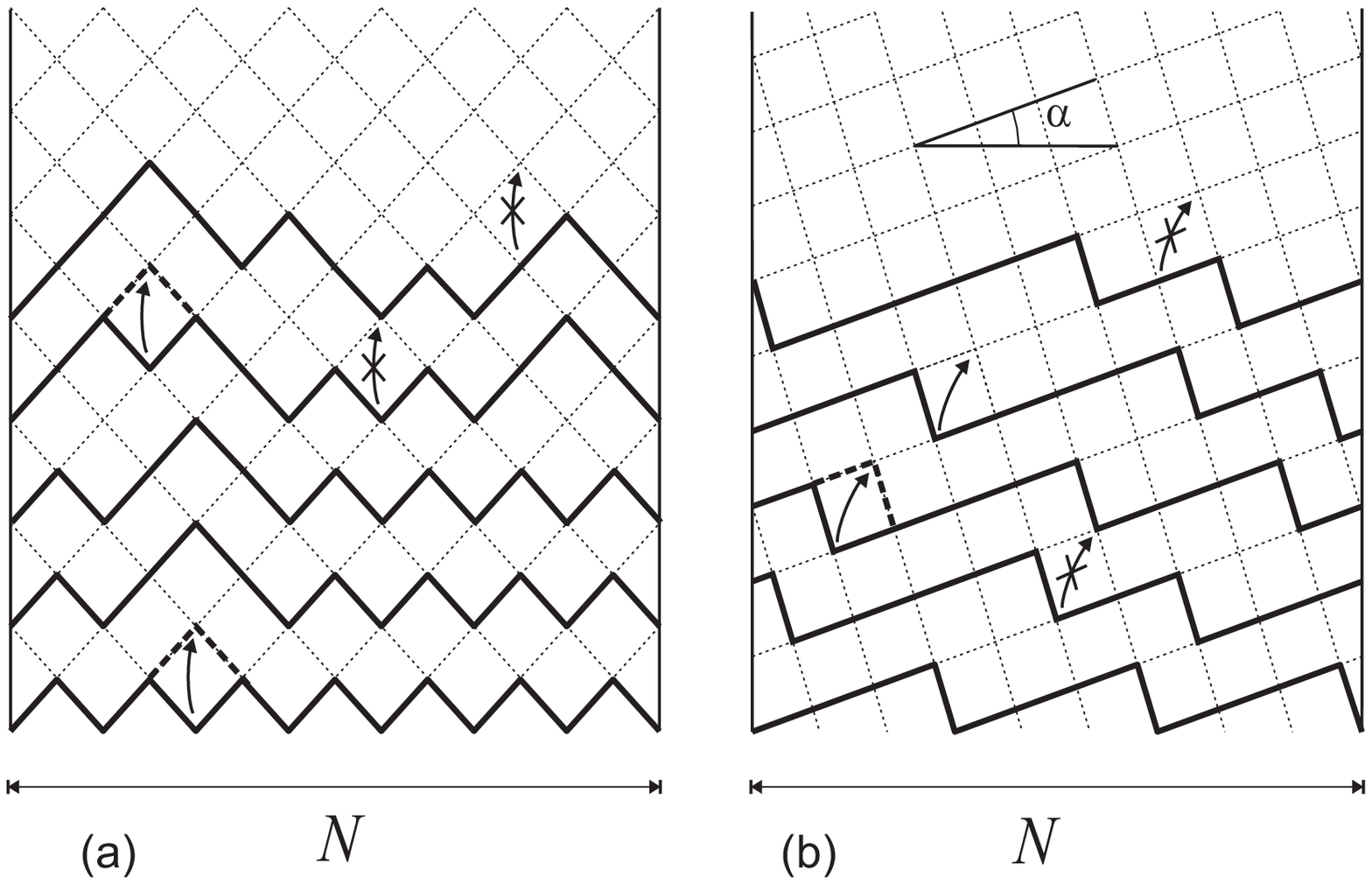,width=7cm}
\caption{The zigzags on a cylinder for different tilt angles $\alpha = \pi/4$ (a) and
$\alpha=\arctan(1/3)$ (b), the examples of allowed an forbidden moves are
shown.}
\label{fig:1}
\end{figure}

Take an infinite cylinder covered by tilted square grid as shown in \fig{fig:1}a,b and consider a
directed closed path (``zigzag'') around a cylinder. Any such path consists of a constant number of
rises and descents, constituting ``kinks'' (here and below we conventionally define rises and
descents with respect to a rightward step). The density of descents, $\rho$, is defined by the tilt
angle $\alpha$ -- see \fig{fig:1}a,b where this density equals $\rho=1/2$ (for $\tan \alpha =1$)
and $\rho=1/4$ (for $\tan \alpha =1/3$), respectively. Consider evolution of a system of such
nonintersecting zigzags. At each infinitesimal time step $dt$ any elementary kink oriented
downwards, can turn upwards with probability $p\,dt$ under the condition that such a move is not
blocked by the upper nearest neighboring zigzag (i.e. all zigzags stay nonintersecting at all
times). By an appropriate rescaling of time, $t$, we set $p\equiv 1$. The examples of elementary
jumps which are allowed and those which are blocked are shown in \fig{fig:1}.

\begin{figure}[ht]
\epsfig{file=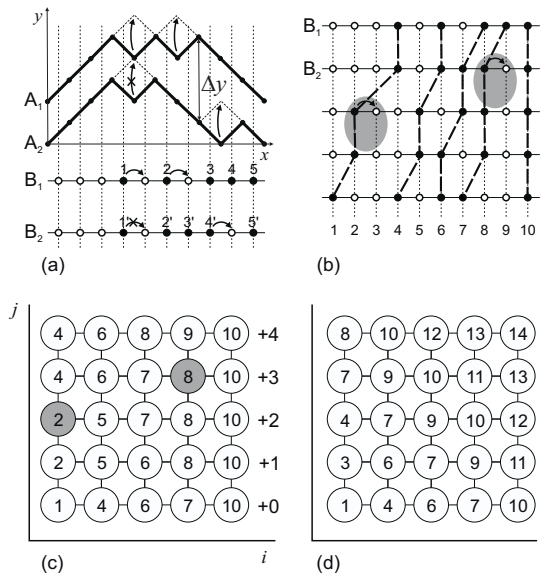,width=7cm}
\caption{The interconnections between zigzag, interacting TASEP
and tag diffusion representations of the model. a) Interaction of two zigzags and the two
corresponding TASEP layers, examples of the same allowed and forbidden moves are shown in both
representations; b) interacting TASEP layers with examples of allowed moves the bold dashed lines
connect the particles with the same number (see explanation in the text); c) the system of tagged
particles corresponding to the TASEP layers depicted in figure (b), with the same allowed moves
highlighted; d) the result of the shift $h_{i,j} \to h_{i,j}+j$.}
\label{fig:2}
\end{figure}

For better understanding of the dynamics of the model, notice that the evolution of a separated
zigzag, (if, for the time being, we neglect its interaction with the other ones), can be
interpreted as a hopping dynamics in a standard one--dimensional Totally Asymmetric Simple
Exclusion Process (TASEP) \cite{pischke} -- see \fig{fig:2} for the corresponding mapping, which is
both conventional and self--explanatory. Essentially, a descent (rise) in a zigzag is identified
with a particle (hole) in the corresponding TASEP. Therefore, the set of zigzags can be viewed as a
system of interacting TASEP layers.

The connection between zigzags and TASEP layers is shown in \fig{fig:2}a. Here, there are two sorts
of constraints on the movement of the particles in layer $B_2$. Indeed, for a jump to be possible
at some point of the $A_2$ zigzag, two conditions should be simultaneously fulfilled: i) it should
be a downward kink, and ii) the movement should not be blocked by the upper zigzag. The translation
of the first condition into the TASEP language is convenient: there should be a particle at a given
position in $B_2$ and a void immediately to the right of it. In turn, the second rule (i.e., the
one describing the interaction of the $B_1$ and $B_2$ layers) can be formulated as follows: it is
possible to label the particles in the layers $B_1$ and $B_2$ in a way such that the $k$th particle
in the layer $B_2$ can never surpass the particle with the same label (i.e., the $k$-th one) in the
upper layer $B_1$, giving rise to an ``interlaced TASEP'' picture shown in \fig{fig:2}b
\cite{endnote2}.

To prove these jumping rules, consider the distance $\Delta y$ between the zigzags $A_1$ and $A_2$
(see \fig{fig:2}a). At two adjacent spatial positions it satisfies
\be \Delta y (x) =
\left\{\begin{array}{ll} \Delta y(x-1)+1 & \mbox{there is a
particle at $x-1/2$}
\\ & \mbox{on $B_2$ and a void on $B_1$}
\medskip \\ \Delta y(x-1)-1 & \mbox{there is a void at $x-1/2$}
\\ & \mbox{on $B_2$ and a particle on $B_1$}
\medskip \\ \Delta y(x-1) & \mbox{otherwise} \end{array} \right.
\label{eq:dist1}
\ee
Note, that $\Delta y(x) \geq 1$ at all positions. Therefore, for each particle on the upper line
there is a ``partner'' particle on the lower line that cannot surpass it, and this ``partnership''
is preserved by the elementary moves. Indeed, if there is a particle on the upper line $A_1$ in
position $x-1/2$ (particle positions are shifted by half--step as compared to the positions of the
kinks) and the zigzag--to--zigzag distance at $x$ is $\Delta y (x)$ then each particle on $A_2$
jumping from $x-1/2$ to $x+1/2$ will decrease $\Delta y$ by one, and, therefore, up to $(\Delta
y-1)$ particles can pass this point without being affected, while the particle number $\Delta y$,
which is the desired ``partner'' particle, will get stacked at $x-1/2$. Moreover, assume now that
the upper particle hops from $x-1/2$ to $x+1/2$. For this move to take place, the position $x+1/2$
should be a void before it, thus (compare to Eq.\ref{eq:dist1})
\be
\Delta y (x+1) = \left\{\begin{array}{ll} \Delta y(x)+1 & \mbox{there is a particle at $x+1/2$}
\\ & \mbox{on $B_2$} \medskip \\ \Delta y(x) & \mbox{otherwise} \end{array} \right.
\label{eq:dist}
\ee
and $\Delta y (x+1)$ does not change as a result of the move. In both cases of (\ref{eq:dist}) the
``partner'' particle after the move is still the same: either the distance and the order of
particles do not change, or the distance increases but the $(\Delta y+1)$th particle as viewed from
the new position is the same as $\Delta y$th from the old position.

In case of a finite cylinder with periodic boundary conditions the ``partner'' particle can, in
principle, lag behind by a whole lap. In this case, one should ensure that the two particles do not
interact even if formally they are at the same place, so one should keep track of the ``real''
distances between the particles (i.e., those which correspond to the distances between the kinks on
the original cylinder), not the ''apparent'' modulo $N$ distances (compare to \cite{priezzhev}).

There is now evidence of some spatial symmetry in the model: for each given particle there are
exactly two other particles, one to the right and one (with the same number) ``on top'' of it --
see the dashed lines connecting the particles of the same number in \fig{fig:2}b, which can block
its movement via excluded--volume interaction. To better exploit this symmetry, it is convenient to
reformulate the model, following the logic of the so-called ``tagged particle diffusion''
introduced in \cite{majumdar} in order to show that on a coarse--grained level the ASEP dynamics is
subject to the one--dimensional KPZ equation.

Consider a set of $m$ TASEP layers (zigzags) of length $N$ with $n$ particles within each layer.
Note that instead of enumerating the cites of TASEP layers and marking which particular cites are
filled with particles, one can store the very same information in a different way by enumerating
the particles with two indices $i \in [1,n]$, $j \in [1,m]$, and ascribing to each particle a
`height' $h_{i,j}$ equal to its position on the corresponding layer (as measured from some
arbitrary chosen ``1st cite''). It is clear now that locally the values in the $h_{i,j}$ matrix are
increasing in $i$ direction and non--decreasing in $j$ direction. To make the model completely
symmetric make the transformation (compare \cite{NM}) $h_{i,j} \to h_{i,j}+j$ which (see
\fig{fig:2}c,d) ensures that $h_{i,j}$ is now increasing function in both directions, $i$ and $j$.
Under this transformation the dynamic rules become totally symmetric:
\be
h_{i,j}\to \left\{\begin{array}{ll} h_{i,j}+1 & \mbox{with Prob. $dt$, if}\,
\left\{\begin{array}{l} h_{i+1,j}>h_{i,j}+1 \\ h_{i,j+1}>h_{i,j}+1 \end{array} \right\}
\medskip \\ h_{i,j} & \mbox{otherwise} \end{array} \right.
\label{eq:1a}
\ee
One can interpret these rules as the dynamics of a heap growing over a two--dimensional substrate,
with an additional constraint of heap gradient being not less than 1 in each of the transverse
directions.

To obtain insights about the large scale dynamics it is useful to consider a coarse--grained
hydrodynamic description of the model~\cite{spohn}. To describe the model at a coarse--grained
level we will henceforth use $(x,y)$ as the spatial coordinates instead of $(i,j)$ in the lattice
model. The coarse--grained dynamics can be described by a local `smooth' velocity field $u(x,y,t)$
which is, in fact, nothing but the average (over space--time element $dx\, dy\, dt$) rate of
successful hops $h_{i,j}\rightarrow h_{i,j}+1$, i.e. the average  probability to find
simultaneously $(h_{i+1,j}-h_{i,j})>1$ and $(h_{i,j+1}-h_{i,j})>1$. This value obviously depends on
the average slope (or equivalently on the local densities in both directions in the Zigzag model)
of the surface in both directions. Similar to the one--dimensional case \cite{spohn,kk} we assume
that the velocity field $u(x,y,t)$ depends on the space-time coordinates only through the local
densities, i.e.,
\be
u(x,y,t)\equiv u(\rho_x(x,y,t),\rho_y(x,y,t))
\label{eq:2}
\ee
where $\rho^{-1}_x(x,y,t)=\frac{\partial h(x,y,t)}{\partial x}$ and
$\rho^{-1}_y(x,y,t)=\frac{\partial h(x,y,t)}{\partial y}$ Note that the value $\rho_x$ introduced
here has a meaning of particle density in the corresponding TASEP layer.

Suppose that the slopes $\rho_{x,y}$ are weakly fluctuating around some average values
$\overline{\rho}_{x,y}$, which are determined by the boundary conditions. Introduce the
displacement $d(x,y,t)$ of a particle located at $(x,y)$ from its average position at $t=0$:
\be
d(x,y,t) = h(x,y,t)-\frac{x}{\overline{\rho}_x} - \frac{y}{\overline{\rho}_y}
\label{eq:3}
\ee
Then, in an analogy with \cite{Maj_thesis} the local values of $\rho_{x,y}$ and $d$ are connected
via
\be
\frac{1}{\rho_x(x,y,t)}= \frac{1}{\overline{\rho}_x} + \frac{\partial d(x,y,t)}{\partial x}; \;
\frac{1}{\rho_y(x,y,t)}= \frac{1}{\overline{\rho}_x} + \frac{\partial d(x,y,t)}{\partial y}
\label{eq:4}
\ee
Assuming now $\left|\frac{\partial d}{\partial x}\right| \ll 1$ and $\left|\frac{\partial
d}{\partial y}\right| \ll 1$ one can expand both $\rho_{x,y}$ and $u(\rho_x,\rho_y)$ as power
series in the derivatives of $d$, obtaining up to the second order:
\be
\begin{array}{l}
\disp \rho_x(x,y,t)=\overline{\rho}_x
\left( 1-\overline{\rho}_x \frac{\partial d}{\partial x}
+ \left(\overline{\rho}_x\right)^2 \left(\frac{\partial d}{\partial x}\right)^2 +
...\right) \medskip \\
\disp \rho_y(x,y,t)=\overline{\rho}_y\left(1-\overline{\rho}_y \frac{\partial d}{\partial y} +
\left(\overline{\rho}_y\right)^2 \left(\frac{\partial d}{\partial y}\right)^2 + ... \right)
\end{array}
\label{eq:5_0}
\ee
which when substituted into \eq{eq:3} gives
\be
\begin{array}{lll}
u(\rho_x, \rho_y) & = &  u(\overline{\rho}_x, \overline{\rho}_y) - \disp \overline{\rho}_x^2
u_x(\overline{\rho}_x, \overline{\rho}_y) \frac{\partial d}{\partial x}  - \overline{\rho}_y^2
u_y(\overline{\rho}_x, \overline{\rho}_y) \frac{\partial d}{\partial y} \medskip \\ & + & \disp
\left( \overline{\rho}_x^3 u_x(\overline{\rho}_x, \overline{\rho}_y)+ \frac{1}{2}
\overline{\rho}_x^4 u_{xx}(\overline{\rho}_x, \overline{\rho}_y)\right)
\left( \frac{\partial d}{\partial x} \medskip \right)^2 \medskip \\
& + & \disp \left(\overline{\rho}_y^3 u_y((\overline{\rho}_x, \overline{\rho}_y) + \frac{1}{2}
\overline{\rho}_y^4 u_{yy}(\overline{\rho}_x, \overline{\rho}_y)\right) \left( \frac{\partial
d}{\partial y} \medskip \right)^2 \medskip \\
& + & \disp \overline{\rho}_x^2 \overline{\rho}_y^2 u_{xy}(\overline{\rho}_x, \overline{\rho}_y)
\frac{\partial d}{\partial x} \frac{\partial d}{\partial y}+...
\end{array}
\label{eq:5} \ee where for brevity we introduced the notation
$u_x$, $u_y$, {\em etc.} for the partial derivatives $u_x =
\frac{\partial u}{\partial \rho_x},\; u_y= \frac{\partial
u}{\partial \rho_y}, ...$.

This allows us to write down a time--dependent differential equation for $d$ which belongs to the
two--dimensional KPZ class:
\be
\begin{array}{lll}
\frac{\partial d}{\partial t} & = & D \Delta d + \eta + \disp u(\rho_x,\rho_y) =
D \Delta d + \eta +\disp u(\overline{\rho}_x,\overline{\rho}_y) -
\medskip \\ & - & \disp \sum_{\alpha=x,y} a_{\alpha}\, \frac{\partial d}{\partial \alpha} +
\sum_{\alpha=x,y} \sum_{\beta=x,y}\, b_{\alpha \beta}\, \frac{\partial d}{\partial \alpha}\,
\frac{\partial d}{\partial \beta}
\end{array}
\label{eq:kpz}
\ee
where $D$ is a diffusion coefficient, $\eta(x,y,t)$ is the noise term
\be
\la \eta(x,y,t) \ra=0,\; \la \eta(x,y,t) \eta(x',y',t') \ra = \delta_{x,x'}
\delta_{y,y'}\delta_{t,t'}
\ee
and $a_{\alpha}, \, b_{\alpha \beta}$  for $\{\alpha, \beta\} = x,y$ are:
\be
\begin{array}{lll}
a_{\alpha} & = & \overline{\rho}_{\alpha}^2 u_{\alpha}(\overline{\rho}_x, \overline{\rho}_y), \medskip \\
b_{\alpha \alpha} & = & \overline{\rho}_{\alpha}^3 u_{\alpha}(\overline{\rho}_x, \overline{\rho}_y)
+ \frac{1}{2} \overline{\rho}_{\alpha}^4 u_{\alpha \alpha}(\overline{\rho}_x, \overline{\rho}_y), \medskip \\
b_{xy} & = & \overline{\rho}_x^2 \overline{\rho}_y^2 u_{xy}(\overline{\rho}_x, \overline{\rho}_y)
\end{array}
\label{eq:coeff}
\ee

\begin{widetext}

\begin{figure}[ht]
\epsfig{file=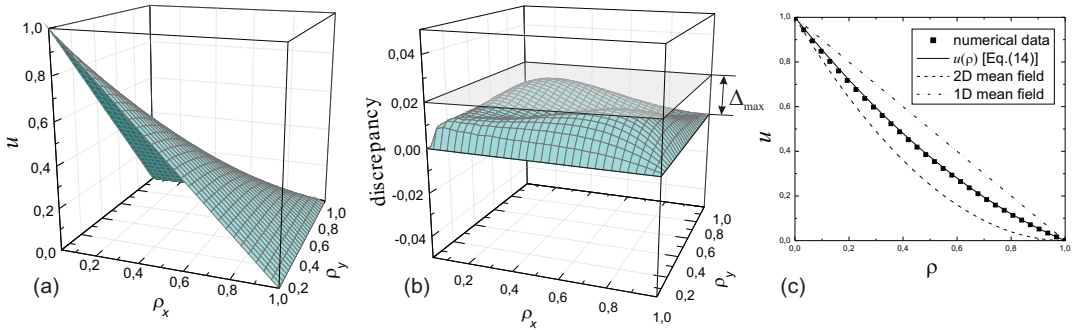,width=14cm} \caption{Comparison of the
theoretical and numerical results for $u(\rho_x,\rho_y)$.
a)$u(\rho_x,\rho_y)$ dependence as measured by direct numerical
simulations (see main text for the details about the simulations);
b) the discrepancy between the numerical result for $u$ and the
one conjectured in Eq. \ref{eq:guess}; c) the $u(\rho)$ behavior
in the $\rho=\rho_x=\rho_y$ plane, the points correspond to
numerical results, while the dashed line is the naive mean-field
guess $u=(1-\rho)^2$ (compare Eq. \ref{eq:meanfield}) and bold
line is the conjectured dependence $u=(1-\rho)(1-\rho/2)$ (compare
Eq. \ref{eq:guess}); .} \label{fig:4}
\end{figure}

\end{widetext}

The behavior of the system is controlled by the usual flow--density dependence $u(\rho_x,\rho_y)$,
which is central in all models of traffic (see, for example, \cite{CSS}) and is usually referred to
as the `fundamental' diagram. Indeed, the flow of the particles in the interlaced TASEP formulation
of the zigzag model is equal to $I(\rho_x,\rho_y)=\rho_x u(\rho_x,\rho_y)$. Recall that in the
1D--case the function $u(\rho)=I(\rho)/\rho$ is just $u=1-\rho$ and it can be obtained by the
mean--field arguments. In absence of interaction between layers one would have expected the same
dependence in our model:
\be
u(\rho_x,\rho_y)=1-\rho_x
\label{eq:1d}
\ee
In the presence of interaction the corresponding 2D mean--field result would be
\be
u(\rho_x,\rho_y)=(1-\rho_x)(1-\rho_y)
\label{eq:meanfield}
\ee
where the symmetry of the model is taken into account: a particle can hop if there is a void both
in front (with probability $1-\rho_x$) and on top (with probability $1-\rho_y$) of it and the
horizontal and vertical jumps are supposed to be independent. However, contrary to the 1D case,
this result is not exact. Indeed, the connectivity of the surface dictates that the local
increments $(h_{i+1,j}-h_{i,j)}$ and $(h_{i,j+1}-h_{i,j})$ are positively correlated and cannot be
considered as independent.

The exact analytical evaluation of the velocity $u(\rho_x,\rho_y)$ up to now is beyond our reach.
We have made numerical simulations of $u(\rho_x,\rho_y)$, the results being presented in
\fig{fig:4}. The simulations were performed for the systems of size $N=32$ in both directions with
periodic boundary conditions ensuring that the average densities take the values $\overline{\rho}_x
=\overline{\rho}_y=0,\frac{1}{31},\frac{2}{31},...,1$ \cite{endnote}. The results were averaged
over $9\times 10^6$ Monte--Carlo steps. To make the comparison with the mean field more visually
compelling, we plot in \fig{fig:4}c separately the numerical data for $\rho_x=\rho_y=\rho$ together
with the mean--field results \eq{eq:1d} and \eq{eq:meanfield}. As expected, the first of them
overestimates the flow, while the second one underestimates it. In fact, the numerical data fits
perfectly the form \be u(\rho)=(1-\rho)(1-\rho/2) \label{eq:guess1} \ee and the consideration of
limiting cases $\rho \rightarrow 0$ and $\rho \rightarrow 1$ suggests that this result may be
exact. In particular by developing the perturbation theory at high densities, we are able prove
that $u(\rho)\to (1-\rho)/2$ as $\rho \rightarrow 1$, the details of these computations will be
published separately \cite{preparation}.

The simplest generalization of Eq.\eq{eq:guess1} onto the case of $\rho_x \neq \rho_y$ which
respects the boundary conditions $u(0,0)=1,\,u(1,\xi)=u(\xi,1)=0$ for any $\xi \in [0,1]$ reads
\be
u(\rho_x,\rho_y)=(1-\rho_x)(1-\rho_y)\left(1+\frac{2 \rho_x \rho_y}
{(\rho_x+\rho_y)(2-\rho_x-\rho_y)}\right)
\label{eq:guess}
\ee
In the \fig{fig:4}b we have plotted the discrepancy $\Delta
u(\rho_x,\rho_y)=u_{\rm n}(\rho_x,\rho_y) - u(\rho_x,\rho_y)$
between numerical, $u_{\rm n}(\rho_x,\rho_y)$ and conjectured,
$u(\rho_x,\rho_y)$, (see \eq{eq:guess}) functions. One sees that
$u(\rho_x,\rho_y)$ is in very good agreement with the results of
numerical simulations.

\begin{figure}[ht]
\epsfig{file=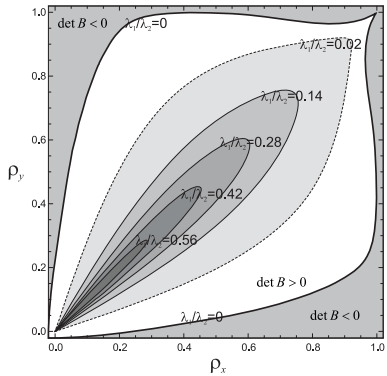, width=4cm}
\caption{Density plot of the ratio $\lambda_1/\lambda_2$ of the eigenvalues of the matrix $B$. In the
region $\det B<0$ the eigenvalues have opposite signs. Outside the dashed area the system can be
regarded as effectively one--dimensional since $\lambda_1 \gg \lambda_2$.}
\label{fig:5}
\end{figure}

The conjectured function $u(\rho_x,\rho_y)$ allows to evaluate the coefficients in \eq{eq:coeff}
for any $\rho_x$ and $\rho_y$ and to calculate the eigenvalues of the coefficients matrix
$b_{\alpha \beta}$ in front of the nonlinear term in Eq.\eq{eq:kpz}. We show thus (see \fig{fig:5})
that the domain $[0<\rho_x<1,\, 0<\rho_y<1]$ can be separated into two regions: i) a region where
one of the eigenvalues dominates ($|\lambda_1|\gg |\lambda_2|$) signaling a quasi--one--dimensional
behavior, and ii) a region where $|\lambda_1|\sim |\lambda_2|$ corresponding to the truly
two--dimensional KPZ dynamics. In the latter region both eigenvalues are negative, and thus the
matrix is positively defined.

Summing up, we have presented a novel model of a statistical driven system in 2+1 dimensions which
turned out to be a direct generalization of the conventional one dimensional TASEP model. In the
hydrodynamic limit we derived the differential equation for the particle displacement in this
model, and we have made a conjecture concerning the flow--on--density dependence of the model (the
so called `fundamental' diagram). Clearly, the model suggested here deserves further investigations
in various directions. In particular, a detailed investigation of the limiting cases $\rho_x=\rho_y
\rightarrow 0$ and $\rho_x=\rho_y \rightarrow 1$ can give valuable understanding of the
flow--density dependence, the study of the displacement fluctuations in different regimes (compare
to \cite{majumdar}) should be very fruitful, and the problem of proving conjecture \eq{eq:guess} or
at least \eq{eq:guess1} sounds quite challenging. Some progresses in these direction will be
reported in a forthcoming longer paper\cite{preparation}.

We are grateful to D. Dhar who have coined the idea of the zigzag model in a private discussion
with one of us (S.M). It is a pleasure for M.T. to express his gratitude for the hospitality of the
LPTMS, Orsay, where the most part of the work has been done.


\begin{thebibliography}{99}
\bibitem{endnote2} Note that the described system of interlaced TASEPs bears
many similarities with one proposed by Borodin and Ferrari in \cite{borodin1,borodin2,borodin3},
ith the exceptiona that there is no ``push--TASEP'' regime in our model.
\bibitem{endnote}  In terms of the original zigzag model our choice of boundary conditions
corresponds to consideration of the system on a torus so that each zigzag consists of $N$ steps,
and there is a room for at most $N$ zigzags on the torus. The densities in the system of $m$
zigzags each of those has $n$ descents is defined  by $\rho_y=(m-1)/(N-1)$; $\rho_x=(n-1)/(N-1)$
(i.e., $\rho_{x}$ and $\rho_y$ are the average densities in the vicinity of a particle excluding
the influence of the particle itself, compare the exact solution for finite TASEP ring).
\end{thebibliography}
\end{document}